# Transport Properties of Heavy Fermion Compounds

D. Jaccard, E. Vargoz, K. Alami-Yadri, and H. Wilhelm

*DPMC, University of Geneva, 24 Quai E.-Ansermet, 1211 Genève 4, Switzerland*

A technique for measuring the electrical resistivity and absolute thermopower is presented for pressures up to 30 GPa, temperatures down to 25 mK and magnetic fields up to 10 T. With the examples of $CeCu_2Ge_2$ and $CeCu_2Si_2$ we focus on the interplay of normal phase and superconducting properties. With increasing pressure, the behaviour of $CeCu_2Ge_2$ evolves from that of an antiferromagnetically ordered Kondo system to that characteristic of an intermediate valence compound as the Kondo temperature increases by about two orders of magnitude. In the pressure window 8-10 < P < 20 GPa, a superconducting phase occurs which competes at low pressure with magnetic ordering. For $CeCu_2Si_2$ the effective mass of carriers is probed by both the coefficient of the Fermi liquid law and the initial slope of the upper critical field. The magnetic instability is studied notably for $CeRu_2Ge_2$ and Yb-based compounds for which pressure-induced magnetic ordering tends to develop. Finally, contrary to conventional wisdom, we argue that in heavy fermions a large part of the residual resistivity is most likely not independent of temperature; tentatively ascribed to Kondo hole, it can be very pressure as well as sample dependent.
[electrical resistivity, thermoelectric power, heavy fermion, magnetic order, superconductivity]

## 1. Introduction

Heavy fermion (HF) compounds are electronic systems close to their magnetic instability. Qualitatively, this characteristic results from the competition of Kondo-like and indirect magnetic RKKY electronic couplings [1]. It is well-known that the former mechanism tends to screen the localised magnetic moments by Kondo spin compensation, while the latter favours magnetic ordering. Moreover, the Kondo and RKKY couplings depend differently on the exchange interaction J between the local f-moment and the conduction electrons and so the ground state of a given system can often be controlled by the application of an external pressure P, since J is a function of P. For instance, it has been shown that a pressure of about 17 GPa can induce HF behaviour in $CeAu_2Si_2$ [2], a compound in which the Kondo effect plays a minor role at P = 0 (small J), while HF systems like $CeCu_2Si_2$ [3] can be driven into the strongly intermediate valence regime (large J) by applying a pressure of 20 GPa. Thus, with an external pressure, the exchange interaction J can be tuned, in particular close to the critical value $J_c$ corresponding to the magnetic/non-magnetic transition observed at a pressure $P_c$.

Transport properties which are very sensitive to the HF phenomenon can be simultaneously measured up to very high pressure and down to very low temperature where the most interesting HF physics develops. Furthermore, a magnetic field can be applied. Transport measurements may offer valuable information about some of the following challenging problems : i) the magnetic phase diagram, especially near the critical pressure $P_c$, ii) the non-Fermi liquid behaviour expected near the magnetic instability, iii) the role of the disorder and iv) the occurrence of superconductivity observed near $J_c$ for an increasing number of Ce-based HF compounds.

The experimental technique to measure the electrical resistivity and absolute thermopower at high pressure is presented in part 2. In part 3, we choose $CeCu_2Ge_2$ from the large $CeM_2X_2$ family where M is a transition metal and X = Si or Ge. For this compound the critical exchange $J_c$ roughly occurs in the middle of the accessible pressure range. Part 4 is devoted to the prototype HF superconductor $CeCu_2Si_2$, while preliminary results for $CeRu_2Ge_2$ are presented in part 5. Finally, the role of the residual resistivity in some Yb-based compounds is discussed in part 6.

## 2. Experiment

We use a simple 'clamp' device derived from the Bridgman technique and from the pioneering work of Wittig [4] to produce pressure. Basically, at room temperature, a load is applied by an oil press on two opposing anvils made out of a hard material. The 'clamp' itself can be easily transferred into a cryostat. In our system, the load can be maintained by four bolts with nuts. The squeezing of the four nuts is easy because the device is suspended by its bolts to stress them during the loading. In order to keep the load (i.e. the pressure) constant at cooling each anvil is supported by a stainless steel piece and the bolts are made out of a titanium-aluminium-vanadium alloy. This choice allows us to balance, within a few percent, the different thermal contractions of the materials used, because the contraction coefficient of the titanium alloy is between that of the anvils and stainless steel supports. Furthermore, the cross-section of the bolts is minimised to reach almost the elastic limit of the titanium alloy when the load is maximum. Additionally, elastic rings increase the elongation at low load. With non-magnetic tungsten carbide (WC) anvils, pressure is limited to about 10 GPa, but the advantage is that an external magnetic field can be applied as the magnetisation of the complete device is sufficiently low. Sintered diamond anvils (Syndie from De Beers) allow higher pressures up to about 30 GPa but are unfortunately magnetic.

Between the anvil flats, the high-pressure cell is placed. A pyrophyllite ring serves as gasket. Its initial internal diameter $\phi$ is between 0.6 and 2 mm depending on the anvil flat, whose diameter is about $2\phi$. The initial thickness is h = 0.125$\phi$. The gasket is glued on one anvil with a sodium silicate solution and is supported by an external epoxy ring which helps the initial formation of the gasket. Inside the gasket, the sample is inserted between two steatite disks used as a pressure transmitting medium. Their diameter is closely adjusted to $\phi$, and their thickness is taken so that the initial filling coefficient of the pyrophyllite ring amounts to 74 %. This ensures the best stability of the gasket. The diameter of the cell will decrease by a few % at low pressure P and almost recover its initial value at high P. This point is essential to limit the development of cracks in intermetallic samples which are often brittle and to guarantee a reliable estimation of the geometrical factor for the resistivity measurements. The samples are obtained by very slow cutting with a thin diamond saw (thickness 25 µm), or by polishing in the case of very brittle compounds. They have a well defined cross-section of typically $25 \times 80$ µm$^2$ and a length of 0.5-1.2 mm. In the high pressure cell, the sample is connected in series with a small Pb-foil whose superconducting transition temperature $T_c$ yields the pressure [5].

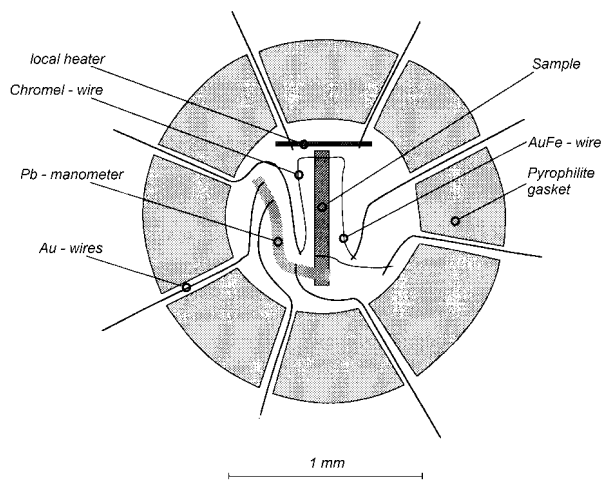

Fig. 1. Schematic drawing of a pressure cell for resistivity and thermopower measurements.

We usually place eight Au or Pt-wires (annealed) across the gasket at about half of its thickness h (See Fig. 1). The wires are wedged in small gorges cut with a razor blade in the gasket and covered with compressed pyrophyllite powder. A wire diameter of about h/5 is a good compromise between the wires breaking and ensuring electrical insulation when the gasket moves. Eight wires allow different possibilities for transport measurements. For instance, the 4-point resistances of two samples in series with the Pb-manometer can be determined. In another configuration, two different parts of the manometer and two different parts of an unique sample can be studied. The former set-up allows, for example, one to compare the behaviour of two samples in almost identical experimental conditions. The latter provides hints on the pressure gradient in the steatite medium and on a possible discrepancy in the pressure response across the sample on a length scale down to about 100 µm i.e. the minimum distance between two voltage wires. The estimation of the pressure gradient is a delicate problem as it changes from cell to cell depending primarily on the stability of the gasket. The width of the superconducting transition of the Pb-manometer gives a local indication of the pressure gradient in the cell. If the pressure is given as $P \pm \Delta P$ then the pressure gradient $\Delta P$ increases slowly with pressure from about 0.1 GPa at P = 1 GPa to 2 GPa at P = 30 GPa.

This kind of high pressure cell also allows us to measure the absolute thermopower. As shown in Fig. 1, in this case the sample in the middle of the cell is carefully aligned with respect to the small local heater made of a Chromel strip. On the end close to the heater, two 12 µm thermocouple wires (AuFe/Chromel) are placed, ideally at exactly the same distance from the heater. With an excitation current of up to 50 mA through the local heater a temperature gradient is obtained. This results in a temperature $T_0 + \Delta T$ at the position of the thermocouple wires ($\Delta T \sim 50$ mK). Tests showed that the other end of the sample can be considered to be sufficiently far away from the heater to remain at $T_0$ because $\Delta T$ decreases roughly as exp(-L/h) where L is the distance to the heater. For this reason one can choose Au for the additional wire used as reference for thermopower measurements and connect the thermocouple wires to 25 µm Au-wires to cross the gasket. The two voltages measured − $V_{AuFe}$ between the AuFe thermocouple and the common Au-wire and $V_{Chr}$ between the Chromel-wire and the Au-wire − let us determine the absolute thermopower $S_X$ of the sample at $T_0 + \Delta T/2$ as $S_X = S_{AuFe} + (S_{Chr} - S_{AuFe})/\{1 - (V_{Chr}/V_{AuFe})\}$, where $S_{AuFe}$ and $S_{chr}$ are the absolute thermopowers of the AuFe and Chromel wires respectively.

The total error in the obtained thermopower (about 15 % at 10 GPa) has several origins. Clearly, a systematic error occurs, whose magnitude depends strongly on the accuracy of the positioning of the thermocouple wires, which should be considered to be of the order of the wire's diameter (12 µm). Indeed, if $\Delta T \propto$ exp(-L/h), one has that $\Delta T$ decreases by about 13 % on a distance L of 12 µm as h ≈ 90 µm for a pressure chamber of internal diameter $\phi = 1$mm. Therefore, a small difference in the distance of the two thermocouples from the heater may lead to a considerable temperature difference. On the other hand, the absolute thermopower of the AuFe-wire and the Chromel-wire are taken to be invariant with pressure, which is, at least in the case of AuFe, a rough approximation. Taking into account these errors, one has to be careful in the interpretation of small variations of the thermopower. Nevertheless the giant thermopower observed in HF systems allows us to get substantial information about these compounds even at very high pressure. Finally we would like to mention that the high pressure cell can be examined after the release of pressure. Putting a drop of alcohol on the steatite makes it transparent. An upper bound for the change of the geometrical factor of the sample, or for the shift of thermocouples can be estimated and correction factors may be introduced.

For cooling our high pressure device down to very low temperature in a dilution refrigerator, special care has to be taken. As the self-clamp system is composed of poor thermal conductors (e.g. stainless steel, Ti-alloy, WC) the thermal path between the mixing chamber of the dilution refrigerator and the WC or sintered diamond anvils is made out of copper pieces. A RuO$_2$ thermometer is fixed on one of the two copper rings enclosing the anvils and is well coupled to the sample, as reflected by the short-time response of the sample to heating. In the case of CeCu$_2$ whose resistivity at 6 GPa exhibits a non-saturating (i.e. likely superconducting) drop below 0.18 K, we were able to show that the whole system, and in particular the measured sample, can be cooled down to 24 mK [6].

### 3. CeCu$_2$Ge$_2$

For CeCu$_2$Ge$_2$ the Kondo effect is already well developed at ambient pressure. However, the RKKY magnetic interaction is still dominant and antiferromagnetic ordering occurs at 4.1 K. Previous work showed that in the vicinity of the pressure-induced magnetic instability, superconductivity emerges at $T_c \sim 0.6$ K [7,8].

The resistivity $\rho$ vs. T of $CeCu_2Ge_2$ at 9.7 GPa is shown in the inset of Fig. 2. Four characteristic temperatures are defined by arrows on this exemplary curve. The resistivity maxima at $T_1^{max}$ and $T_2^{max}$ reflect the Kondo effect on the ground and excited crystal field levels, respectively. At low temperature, the kink in $\rho(T)$ at $T_M$ indicates magnetic ordering and at $T_c$ the sudden drop of $\rho$ to zero ($\rho < 10^{-4}$ $\mu\Omega cm$) marks the superconducting transition. The P-dependence of these four characteristic temperatures is depicted in Fig. 2. The rapid rise of $T_1^{max}$ above ~ 7 GPa indicates the increase of the Kondo coupling. For $P \sim 17$ GPa, when the system approaches the intermediate valence regime, $T_1^{max}$ and $T_2^{max}$ merge, and only one resistivity maximum is observed as for the thermopower [9]. At low pressure, two magnetic ordering temperatures $T_M$ seem to occur in qualitative agreement with [10]. For intermediate P, the disappearance of magnetic ordering corresponds to the emergence of superconductivity apparently with a narrow pressure window where both phenomena coexist. The critical temperature $T_c \sim 0.6$ K remains constant up to 13 GPa. At higher P, $T_c$ increases rapidly, reaches almost 2 K at $P \sim 16$ GPa and then vanishes at $P \sim 20$ GPa. Due to a small pressure gradient in the measuring cell, the width of the transition $\Delta T_c = T_c^{10\%} - T_c^{90\%}$ increases when $T_c(P)$ varies rapidly.

Fitting the Fermi liquid relation $\rho = \rho_0 + AT^2$ to our data up to a temperature $T_A$, indicates that the $T_c(P)$ maximum is closely related to anomalies in the P-dependence of both the residual resistivity $\rho_0$ and coefficient A. Indeed, as shown in Fig.3, a pronounced peak in $\rho_0$ and a shoulder in A occur at almost the same pressure as the $T_c(P)$ maximum. Moreover, in agreement with usual scaling, one has approximately $A \propto T_A^{-2}$ and also $A \propto (T_1^{max})^{-2}$ which supports the correlation of the $T_1^{max}$ with $T_K$. For the effective masses this would mean a drop by a factor ~50 at 25 GPa in comparison to the maximum values expected at the magnetic instability. Obviously the strong features observed at 15-16 GPa can be considered as signs of a phase transition (e.g. Kondo collapse).

The P-dependence of $T_c$ is very similar to what was observed for $CeCu_2Si_2$ at pressures about 10 GPa lower. In fact all the measured properties of $CeCu_2Ge_2$ and $CeCu_2Si_2$ are almost identical if this pressure shift is taken into account. A comparison with previous work [7,8] shows that the pressure $P_c$ at which the magnetic ordering disappears is somewhat sample dependent suggesting stoichiometric effects as for $CeCu_2Si_2$ at $P = 0$. One important point is that the data of Fig. 2 and 3 have been measured in a 8-wire cell (see part 1) and correspond to one part of a single crystalline sample. The other part behaves quite similarly but the above-mentioned characteristic temperatures and values of $\rho_0$ and A are shifted by about 1.5 GPa towards lower P, i.e. in excess of the estimation of the pressure gradient near $P_c$. One then has evidence that the pressure response of the sample varies substantially on a submillimetric scale. All happens as if mean values were observed and any discussion of, for example, the coexistence of superconductivity and magnetism in $CeCu_2Ge_2$ seems perilous.

The Fermi liquid law for $\rho$ is the most reliable fit to our low T data for all $P > P_c$. Above 16 GPa, the $T^2$ law is clear in large T-windows with sizeable coefficients A. At lower P, the fit is less obvious because the T-window between $T_c$ and $T_A$ shrinks. Previously, just above $P_c$, the $T^2$ dependences for $0.1 < T <\sim 1$ K have been found in a magnetic field (in order to suppress superconductivity) [7,8]. For intermediate P, the apparent linear T-behaviour of $\rho$ (up to about 15 K i.e. ~ 5 $T_A$ for $P = 16$ GPa) [11] is an artefact due to the change of the $\rho(T)$ curvature and to the interplay with the residual term. The results for $CeCu_2Si_2$ will improve our understanding of this intriguing link.

Let us finally mention, that other $CeCu_2Ge_2$ single crystals have been studied. Generally, samples with too high residual resistivity present only a partial resistivity drop at $T_c$ and/or do not show any $T_c$ increase above 12 GPa.

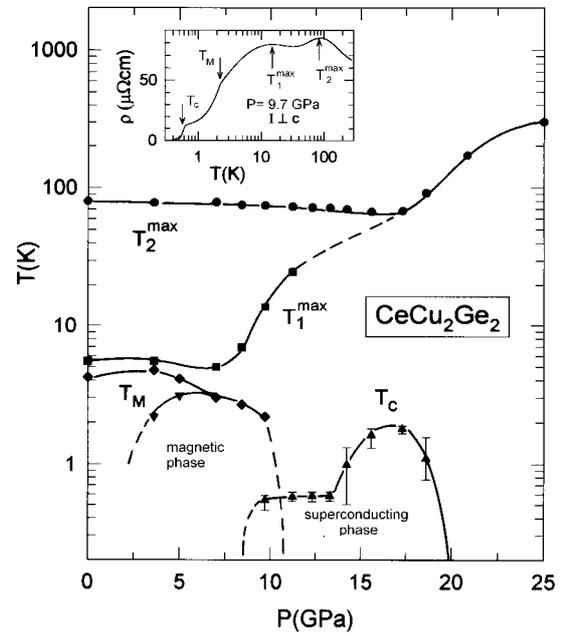

Fig. 2. Pressure dependence of the temperatures of magnetic ordering $T_M$, superconducting transition $T_c$ and maxima of resistivity $T_1^{max}$ and $T_2^{max}$ as defined in the inset, for $CeCu_2Ge_2$. Vertical bars indicate the width of $T_c$. ($\Delta T_c = T_c^{10\%} - T_c^{90\%}$).

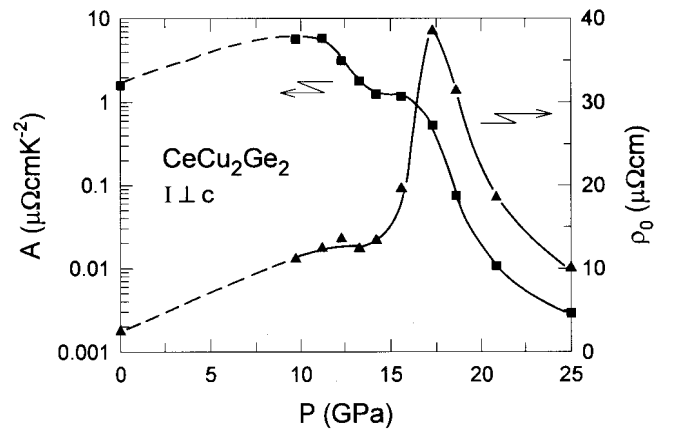

Fig. 3. Pressure dependence of the residual resistivity $\rho_0$ and of the coefficient A in the relation $\rho = \rho_0 + AT^2$, for $CeCu_2Ge_2$.

## 4. CeCu$_2$Si$_2$

Recently two single crystals of CeCu$_2$Si$_2$ (labelled below as sample I and II) have been simultaneously measured (part 2) to study the role of $\rho_0$, to determine the upper critical field H$_{c2}$, and in particular its initial slope. Sample I has been grown by slow cooling under 50 bar Ar atmosphere in a BaZrO$_3$ crucible. At P = 0, the $\rho_0$ and A values were 5 µΩcm and 7 µΩcm/K$^2$. Sample II exhibited a higher $\rho_0$ of 21 µΩcm and a lower A of 4 µΩcm/K$^2$. Both samples showed narrow superconducting transitions ($\Delta T_c$ ~ 20 mK) at 0.74 and 0.64 K, respectively. In a magnetic field, $\Delta T_c$ remained narrow and H$_{c2}$(T) values were in good agreement with the literature [12]. Below 1 K, the magnetoresistance of the normal phase was small and positive. For sample I, the transition at B = 7.5 T between phases A and B [13] coincided with a sudden 15 % drop in $\rho$ at 50 mK.

At ambient pressure, CeCu$_2$Si$_2$ is close to the magnetic instability (P$_c$ ~ 0), and at 6.5 GPa the low and high T maxima of $\rho$ have just merged (compare with CeCu$_2$Ge$_2$ in Fig. 2). Fig. 4 shows $\rho$(T) for sample I and II at P = 6.5 and 9.9 GPa. Sample I shows the expected behaviour, in good agreement with [3]. At 6.5 GPa, $\rho_0$ takes its maximum value of ~ 20 µΩcm and at low T one has $\rho \propto T^2$. For sample II, the role of $\rho_0$ clearly appears. Above ~ 50 K and in comparison with sample I, $\rho$ is simply shifted upwards roughly by a constant value. The resistivities of both samples have maximum values at about the same temperature pointing to similar T$_K$ estimations. The nearly constant shift of $\rho$ may be partly due to systematic errors in the geometrical factor of the samples and can be ascribed to a static residual term which seems to obey the Matthiessen rule. In contrast to sample I, $\rho$(T) of sample II shows an upturn at low temperature, which could be due to an effect of Kondo holes [14] and $\rho_0$ rises up to the very high value of ~ 170 µΩcm. Very similar resistivity behaviour (upturn, T$^{max}$ and magnitude of $\rho$) has been reported for non-stochiometric CePd$_3$ [15].

The inset of Fig. 4 shows that the $\rho$(T) upturn follows a -T$^2$ law which supports an explanation in terms of Kondo hole. Thus it appears that the resistivity has three contributions: i) a static residual term which should be small (~ 1 µΩcm for sample I), ii) an impurity term apparently $\propto$ -T$^2$ and iii) a pure (ideal) lattice term. Even for sample I, the impurity term contributes to $\rho$ and the measured A coefficient is reduced in comparison with the pure lattice value. The impurity and lattice terms are likely to be non additive and then deviation of a quadratic law may happen.

Resistivity data for CeCu$_2$Si$_2$ in a magnetic field are presented in Fig. 5 for sample I at 4.5 GPa. The inset shows the upper critical field H$_{c2}$ at different P taking T$_c^{onset}$ as a measure of T$_c$. At 4.5 GPa T$_c$ is relatively narrow because T$_c$(P) is near its maximum and then does not vary much in the P-window $\Delta$P. We note that the transition is not quite complete and a small ohmic contribution remains down to ~ 1.5 K at B = 0. This temperature is about the maximum value of T$_c^{onset}$ of sample II. One also sees in Fig. 5 that the T$^2$ law of the normal phase is obvious only at low T. In the inset, the extrapolation to T = 0 indicates that qualitatively H$_{c2} \propto$ T$_c$. But the most interesting result is that the initial slope ($\partial H_{c2}/\partial T$)$_{Tc}$ is almost P-invariant and this conclusion does not depend on the T$_c$ criteria (T$_c^{onset}$, T$_c^{10\%}$ or T$_c^{90\%}$) used.

Considering that in clean limit conditions, ($\partial H_{c2}/\partial T$)$_{Tc}$/T$_c$ should scale with the square of the effective mass m*, one deduces that m* has decreased by about a factor of 2 at 4.5 GPa as compared with P = 0. This estimation agrees with the drop of the A coefficient. However, at higher P, A decreases very rapidly (as for CeCu$_2$Ge$_2$ for P > 15 GPa, see Fig. 3) and leads at 6.5 GPa to a much lower m* value than the initial slope H$_{c2}$'. The discrepancy remains even if an under-estimation of A due to an impurity Kondo hole effect is taken into account. Experiments are now in progress to obtain a better determination of H$_{c2}$(T,P) and thus a better reference to discuss the validity of theoretical models (clean or dirty limit, weak or strong coupling).

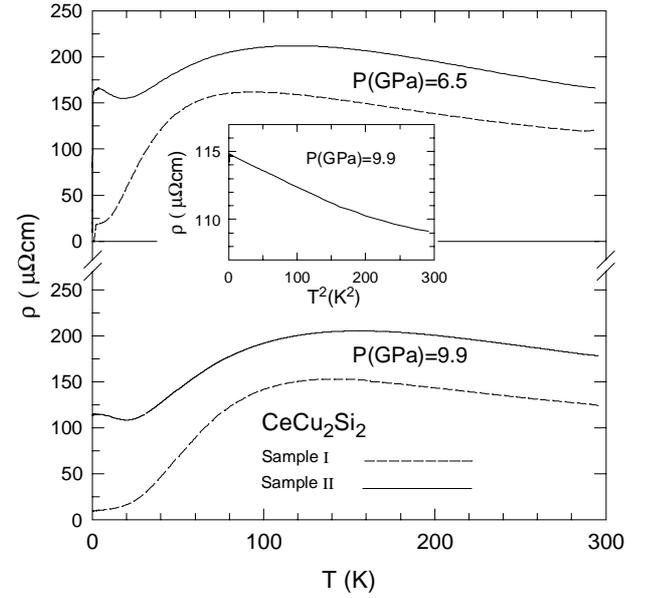

Fig. 4. Resistivity of two different CeCu$_2$Si$_2$ single crystals at 6.5 and 9.9 GPa. Inset: T$^2$ plot at low T for sample I.

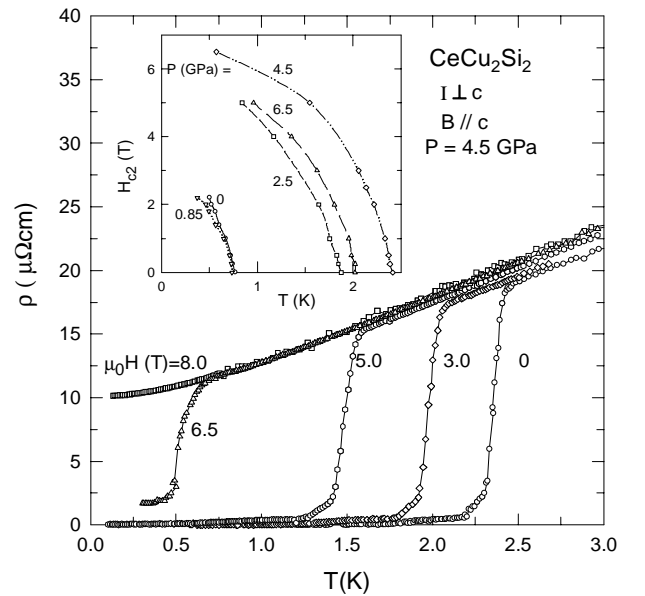

Fig. 5. Low T resistivity of single crystalline CeCu$_2$Si$_2$ at selected magnetic fields. Inset: upper critical field H$_{c2}$' vs. T.

## 5. CeRu$_2$Ge$_2$

The striking aforementioned correspondence of CeCu$_2$Ge$_2$ with CeCu$_2$Si$_2$, whose behaviour appears as simply shifted on a P-scale, should be more general and should work in particular for CeRu$_2$Ge$_2$ and CeRu$_2$Si$_2$. According to the literature the former system has a ferromagnetic ground state and the Kondo effect is negligible [16], while the latter is a non-magnetic HF which exhibits a metamagnetic transition at 8.2 T [17]. A more specific feature of CeRu$_2$Ge$_2$ is the strong competition between ferromagnetic and antiferromagnetic couplings which persists in CeRu$_2$(Ge$_x$Si$_{1-x}$)$_2$ alloys down to the magnetic instability at $x_c \sim 0.1$ [16]. Moreover, experiments and band structure calculations confirm that the Fermi surfaces of the pure Ge (x = 1) and Si (x = 0) compounds enclose a volume which differs by one electron, i.e. the 4f electron of Ce [18].

Preliminary resistivity data of CeRu$_2$Ge$_2$ (Fig. 6) show that the magnetic/non-magnetic transition is located at a pressure $P_c$ between 4.5 and 9.9 GPa. Moreover, the extrapolation of the P-dependence of the ferromagnetic ordering temperature yields $P_c \sim$ 5-6 GPa. In comparison with Cu-based isotype compounds, the scattering at $P_c$ is much lower and the temperature $T^{max}$ of the Kondo resistivity maximum is not clearly resolved. In the inset one sees that the Fermi liquid law is obeyed at 6.3 and 9.9 GPa only up to temperatures $T_A$, of 0.55 and 1.12 K, respectively and with small values of the A coefficient. However, A as well as $\rho_0$ seem to reach their maximum at $P_c$. These results agree well with those obtained for Ce$_{1-x}$La$_x$Ru$_2$Si$_2$ [19], for which the magnetic instability is reached at $x_c \sim 0.1$. The relatively low scattering at $P_c$ points to moderate heavy masses. This characteristic is a common feature with CeRh$_2$Si$_2$ [20] and CePd$_2$Si$_2$ [21] i.e. when a d-transition metal is substituted for Cu. Nevertheless, it could happen that the low resistivity is due to parallel conduction channels of lighter carriers.

The mentioned correspondence of CeRu$_2$Ge$_2$ to CeRu$_2$Si$_2$ is seen in magnetoresistance measurements performed at 0.06, 0.8 and 4.2 K for P = 6.3 GPa. To a large positive magnetoresistance a shoulder at $\sim$ 3 T is superimposed. This feature is almost T-invariant, and is tentatively ascribed to the metamagnetic transition also seen in CeRu$_2$Si$_2$. It seems to vanish at 9.9 GPa.

## 6. Ytterbium-based compounds

Yb-based compounds offer an alternative way of studying the magnetic instability of HF, because the exchange J decreases with P, contrary to the Ce-case. For instance, YbCu$_2$Si$_2$ is an intermediate valence system at P = 0 and its magnetic resistivity shows a single maximum at $T^{max} \approx$ 150 K, as for CeCu$_2$Ge$_2$ at $\sim$ 22 GPa. By contrast, the resistivity of YbCu$_2$Si$_2$ at $\sim$ 22 GPa has regained two maxima [11], as observed for CeCu$_2$Ge$_2$ at P = 0, because $k_B T_K$ has become much smaller than the crystal field energies. Hence more generally, pressure qualitatively acts as a mirror between Ce and Yb compounds. At intermediate P, evidence of magnetic ordering of YbCu$_2$Si$_2$ has already been reported [22]. A small part of a systematic investigation of Yb-based compounds is briefly presented below.

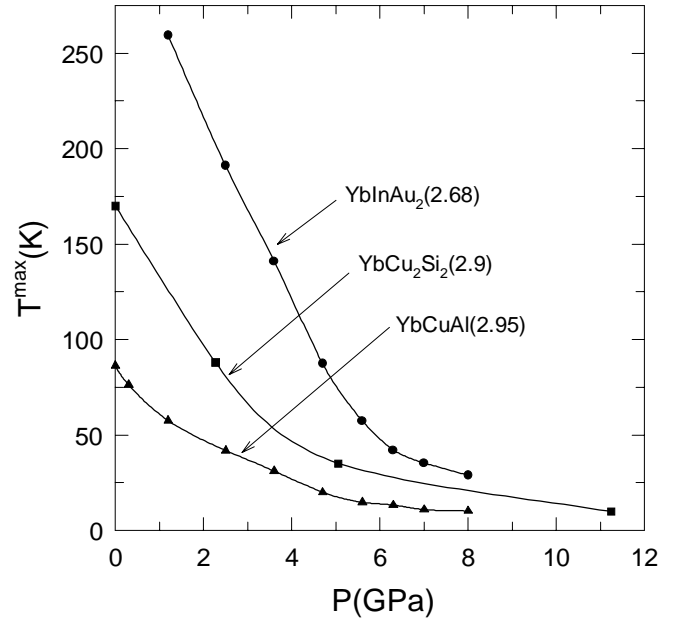

Fig. 7. Temperature $T^{max}$ vs. P of the resistivity maximum of YbInAu$_2$, YbCu$_2$Si$_2$ and YbCuAl. Valence at 300 K estimated by L$_{III}$-edge experiments is indicated.

Fig. 7 shows the P-dependence of the temperature $T^{max}$ of the $\rho$(T) maximum for three Yb-compounds. By comparison with Ce-based compounds (see e.g. the cases of CeAu$_2$Si$_2$ and CeCu$_2$ [11]), the magnetic ordering should appear near the change of the $T^{max}$ slope. However, signs of magnetic ordering are most often unclear, most likely because the residual scattering strongly increases when the system approaches its magnetic instability. For YbCuAl the residual $\rho_0$ rises up to $\sim$ 100 $\mu\Omega$cm and for YbInAu$_2$ low T upturns of $\rho$(T) (although of weak magnitude) are observed near 4 GPa. Such upturns which develop in the P-range where $\rho_0$ increases can be considered as an indication that the Kondo hole contributes to the residual scattering. The case of $YbCu_2Si_2$ is shown in Fig. 8 for different samples. One sees that the magnitude of the residual peak is sample dependent as emphasised by the dark symbols which refer to two different parts of the same sample simultaneously measured (see part 2). Interestingly, the sample (open circle) for which the most convincing signs of magnetic ordering (kink in $\rho(T)$, $T^3$ power law) have been found [22], does not show any marked residual peak. For P > 8 GPa, the other samples show magnetic anomalies below about 3 K which vanish in a field of ~ 3 T. Measurements at 9.2 GPa show that in a field of 8 T, $\rho_0$ recovers the value observed at low P. Accordingly, the residual peak of $YbCu_2Si_2$ seems strongly field dependent, suggesting a low characteristic Kondo hole energy. In contrast, the residual $\rho_0$ of YbCuAl and $YbInAu_2$ is field independent. Let us finally mention that in the P-range where the $\rho(T)$ curvature changes with increasing P from positive to negative values, a power law of the form $\rho \propto T^n$ with n ~ 1 can be found.

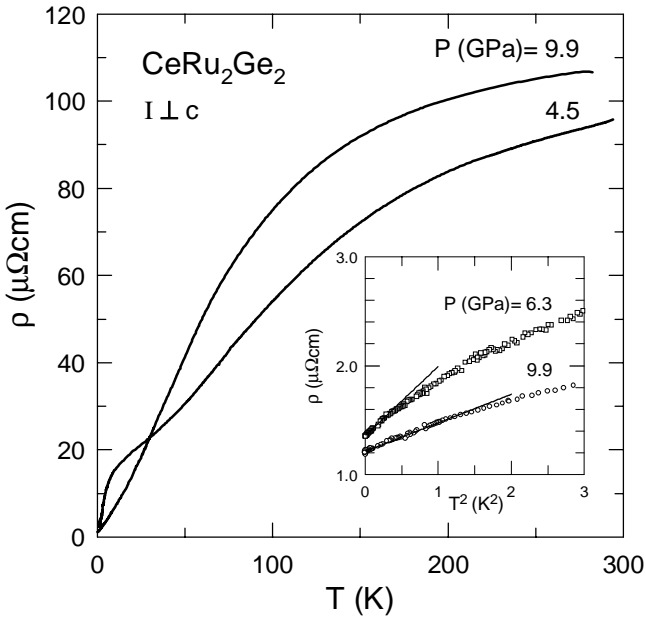

Fig. 6. Resistivity vs. T of CeRu$_2$Ge$_2$ at 4.5 and 9.9 GPa. Inset: $T^2$ plot of the low T data at 6.3 and 9.9 GPa.

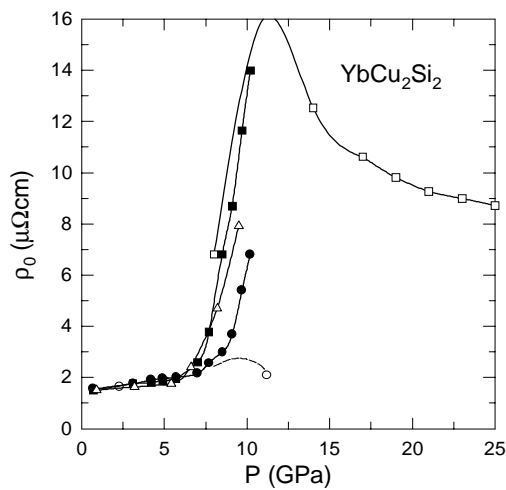

Fig. 8. Residual resistivity vs. P of different $YbCu_2Si_2$ samples. Dark symbols refer to different parts of a same sample.

## 7. Concluding remarks

Detailed studies of the ac susceptibility of $CeCu_2Si_2$, using solid helium as the pressure transmitting medium, have confirmed and shown precisely the $T_c(P)$ dependence [23]. Notably two cusp-like anomalies have been detected at 3 and 7.6 GPa. Pressure-induced topological changes of the Fermi surface have been proposed to explain the $T_c$ increase and the first cusp-like anomaly superimposed on the more usual pressure-induced decrease of $T_c$ for a narrow-band superconductor [23]. In our resistivity experiments, sudden change of the $T_c$ slope cannot be resolved due to the pressure gradient in the cell. The quite similar properties of $CeCu_2Si_2$ and $CeCu_2Ge_2$ appeal for the same interpretation of the $T_c(P)$ maximum. This is furthermore clearly correlated with anomalies in the normal phase and occurs when the system approaches its intermediate valence regime. The challenge is now to incorporate the Kondo effect in the aforementioned $T_c$ scenario and to simultaneously explain the peak of $\rho_0$ and the shoulder of A close to $P_c$. Above the pressure at which $T_c$ is maximum, the rapid drop of A leads to much lower m* values than that deduced from the initial slope $H_{c2}$'. This discrepancy will be investigated in more detail. More generally, one needs to improve the hydrostaticity of our high pressure cell, using helium as the transmitting medium.

There is now a large amount of data (see e.g. the case of $CeAl_3$ [8,24]) which show that the residual resistivity of HF compounds is most often very dependent upon pressure and magnetic field, especially near the magnetic instability. For $CeCu_2Si_2$ and $CeCu_2Ge_2$, the $\rho_0$ peak, observed in coincidence with the $T_c$ maximum, seems to be an exception to this rule as it occurs far above $P_c$. For $CeCu_2Ge_2$ only a plateau of $\rho_0$ is observed around $P_c$. However the example of $YbCu_2Si_2$ emphasises that $\rho_0$ can be extremely sample dependent near $P_c$. Furthermore, for the superconducting compounds, the residual scattering plays a subtle role in the pair breaking mechanism. Finally there is little evidence that the disorder can be only reduced to an independent temperature term $\rho_0$. It is more likely that only one part of $\rho_0$ due to static disorder can be subtracted from the total resistivity according to the Matthiessen rule. As a result, a power law relationship for $\rho$ should be considered with caution even in the lattice case.

**Acknowledgement**. We would like to thank A. Erb and E. Walker for providing us with $BaZrO_3$ crucibles and J.-Y. Genoud for preparing sample I. We are also grateful to B. Lüthi who gave us sample II prepared by W. Assmus. This work was partly supported by the Swiss National Science Foundation.